\documentclass[prd,aps,twocolumn]{revtex4}

\usepackage{revsymb}
\usepackage{graphicx}% Include figure files
\usepackage{dcolumn}% Align table columns on decimal point
\usepackage{bm}% bold math
\usepackage{color}
\usepackage{amsmath,amsthm,amssymb}
\usepackage{url}
\begin{document} 

\title{Longitudinal and transverse spin asymmetries for inclusive jet production at mid-rapidity in polarized $p+p$ collisions at $\sqrt{s}=200$ GeV}
%\vspace{-0.1in}

%Version 24 plus many cag edits: Oct 24, 2011

%\date{\today}
\author{
L.~Adamczyk$^{1}$,
G.~Agakishiev$^{20}$,
M.~M.~Aggarwal$^{31}$,
Z.~Ahammed$^{50}$,
A.~V.~Alakhverdyants$^{20}$,
I.~Alekseev$^{18}$,
J.~Alford$^{21}$,
B.~D.~Anderson$^{21}$,
C.~D.~Anson$^{29}$,
D.~Arkhipkin$^{3}$
% ...
}

\address{$^{1}$AGH University of Science and Technology, Cracow, Poland}
% ...
\address{$^{3}$Brookhaven National Laboratory, Upton, New York 11973, USA}
% ...
\address{$^{18}$Alikhanov Institute for Theoretical and Experimental Physics, Moscow, Russia}
% ...
\address{$^{20}$Joint Institute for Nuclear Research, Dubna, 141 980, Russia}
\address{$^{21}$Kent State University, Kent, Ohio 44242, USA}
% ...
\address{$^{29}$Ohio State University, Columbus, Ohio 43210, USA}
% ...
\address{$^{31}$Panjab University, Chandigarh 160014, India}
% ...
\address{$^{50}$Variable Energy Cyclotron Centre, Kolkata 700064, India}
% ...
\vspace{-1.0cm}
\begin{abstract}
\vspace{-0.5cm}
\begin{center}
(The STAR Collaboration) 
\end{center}
\vspace{-0.3cm}
%\vspace{-0.2cm}
We report STAR measurements of the longitudinal double-spin asymmetry $A_{LL}$, the transverse single-spin 
asymmetry $A_{N}$, and the transverse double-spin asymmetries $A_{\Sigma}$ and $A_{TT}$ for inclusive jet 
production at mid-rapidity in polarized $p+p$ collisions at a center-of-mass energy of $\sqrt{s} = 200\,\mathrm{GeV}$. 
The data represent integrated luminosities of $7.6\,\mathrm{pb}^{-1}$ with longitudinal polarization and 1.8 pb$^{-1}$ 
with transverse polarization, with 50-55\% beam polarization, and were recorded in 2005 and 2006. No evidence is found 
for the existence of statistically significant jet $A_{N}$, $A_{\Sigma}$, or $A_{TT}$ at mid-rapidity. Recent model calculations indicate 
the $A_N$ results may provide new limits on the gluon Sivers distribution in the proton. The asymmetry $A_{LL}$ significantly 
improves the knowledge of gluon polarization in the nucleon.
\end{abstract}

% PACS, the Physics and Astronomy Classification Scheme.
% spin,proton structure, jets, pol., gluons, inel. p scat.,QCD test
%\keywords{jets, gluon, asymmetry, longitudinal, transverse}%Use showkeys class option if keyword
%display desired

\maketitle
%\pagewiselinenumbers
\section{Introduction}

Deep-inelastic scattering (DIS) experiments with polarized lepton beams and targets containing
polarized nucleons have measured the inclusive spin structure function $g_1(x,Q^2)$ of the
nucleon over a wide range in Bjorken-$x$, $0.003 < x < 0.8$, and $Q^2$, $1 < Q^2 < 100$~GeV$^{2} /c^2$ 
\cite{Ashman:1987hv}-\cite{Alekseev:2010hc}.
%(see references in \cite{Blumlein:2010rn,Leader:2010rb}). 
The DIS data, combined with data on the couplings in neutron and
hyperon $\beta$ decay, lead one to conclude that the quark contribution to the spin of a
longitudinally polarized nucleon is only about 25\%, well below naive expectations that the
nucleon spin originates mainly from the valence quarks. Perturbative QCD
analyses \cite{Blumlein:2010rn}-\cite{deFlorian:2009vb}
%\cite{Blumlein:2010rn,Leader:2010rb,deFlorian:2005mw,deFlorian:2009vb}
of the $Q^2$ dependence of $g_1(x,Q^2)$ gave the first insights
into possible gluon polarization contributions, but the precision is thus far limited
by the $Q^2$ range that is accessible in the fixed-target experiments. Semi-inclusive
DIS spin asymmetry measurements with identified pions and kaons
have made it possible to delineate the quark and anti-quark spin contributions by flavor, and
measurements with hadron pairs and open charm mesons have shown sensitivity to gluon
polarization \cite{alek:2009}-\cite{aira:2010um}.

Collisions of polarized proton beams at the Relativistic Heavy Ion Collider (RHIC)
at Brookhaven National Laboratory have made it possible to study proton spin structure
via hadroproduction of jets and other hard probes at ten-fold higher center-of-mass
energies than previous DIS experiments. Of particular interest to the determination
of gluon polarization is the longitudinal double-spin asymmetry $A_{LL}$,
\begin{equation}
A_{LL} = \frac{\sigma^{++} - \sigma^{+-}}
{\sigma^{++} + \sigma^{+-}},
\end{equation} 
\noindent where $\sigma^{++}$ and $\sigma^{+-}$ are the differential production cross 
sections when the beam protons have 
equal and opposite helicities, respectively. STAR has published differential production cross section 
data for inclusive jet production at mid-rapidity with transverse momenta, $p_T$, 
in the range $5 < p_T < 50$~GeV/$c$ \cite{Abelev:2006uq} 
that are well described by perturbative QCD calculations at next-to-leading order 
(NLO) \cite{Jager:2004jh}. 
This supports the use of this framework in interpreting our measurements of $A_{LL}$.

Semi-inclusive deep-inelastic scattering experiments have also measured a broad range of transverse 
spin asymmetries (see for example \cite{Aira2005}-\cite{xqian}), 
%\cite{Aira2005,Alek2009,Aira2009,Alek2010}), 
including asymmetries 
sensitive to the Collins and Sivers effects.  The Collins effect involves the convolution of quark 
transversity with the transverse-spin-dependent Collins fragmentation function, which has been measured 
in $e^{+}e^{-}$ scattering \cite{Abe2006,Seidl2008}.  The Sivers effect ascribes a spin-dependent 
transverse momentum to the partons in a transversely polarized proton.  Recently, global analyses 
have been performed to extract the quark transversity \cite{Ansel_collins09} and parton 
Sivers \cite{Ansel_sivers09} distributions from the semi-inclusive DIS and $e^{+}e^{-}$ data.  
Measurements of the transverse double-spin asymmetry, $A_{TT}$, for inclusive jet production 
provide a complementary probe of quark transversity \cite{Soffer02}.  It has also been proposed 
that the transverse single-spin asymmetry, $A_N$, for inclusive jet production may be sensitive 
to the Sivers effect \cite{DAlesio11}.

In this article, we discuss the techniques that STAR uses to find and reconstruct 
inclusive jets in polarized $p+p$ 
collisions, update our earlier analysis of $A_{LL}$ from 2005 data \cite{Abelev:2007vt}, 
and present precision data recorded in 2006 on $A_{LL}$.
The $A_{LL}$ results significantly improve our knowledge of the gluon polarization in the proton. 
We also present the first results for the transverse single-spin asymmetry $A_{N}$ and 
transverse double-spin asymmetries $A_{\Sigma}$ and $A_{TT}$ (defined in Sect.\@ \ref{sec:trans_spin}) 
for the inclusive production of mid-rapidity jets with 
transverse momenta up to 35~GeV/$c$ in polarized proton collisions at $\sqrt{s} = 200\,\mathrm{GeV}$. 
The $A_N$ measurement may provide new limits on the gluon Sivers distribution in the proton \cite{DAlesio11}.

\section{Experiment and Data} 

\subsection{RHIC-STAR}
\begin{figure}
\includegraphics[width = 0.45\textwidth]{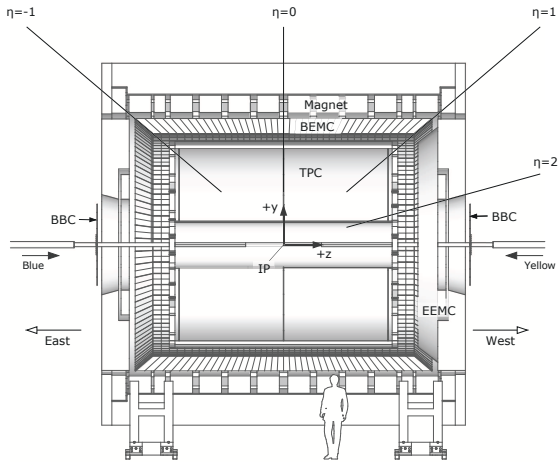}
\caption{\label{star} Schematic section cut of the STAR detector showing the
detector elements used in these measurements.}
\end{figure}

A schematic diagram of the STAR detector at RHIC is shown in Fig.~\ref{star}. 
The detector studies collisions of independently polarized proton beam bunches, 
ranging from 50-112 bunches stored in each of two rings for a given fill. During the 2005 run, the proton 
bunches were loaded 
with alternating spin directions for the beam that circulated in the clockwise direction 
(blue) and with alternating spins for successive pairs of bunches for the beam that circulated 
in the counterclockwise direction (yellow). More complex 8-bunch polarization patterns were 
implemented in 2006 to further reduce possible systematic errors associated with individual bunch 
patterns. Collisions at STAR are tagged by a coincidence of hits in the east and west 
Beam-Beam Counters (BBC)~\cite{BBCref}, scintillation detectors that consist of 18 
hexagonal tiles subtending the pseudorapidity interval, $3.4 < |\eta| < 5.0$, on opposite 
sides of the interaction region. The performance of this detector as a luminosity monitor 
was cross-checked against a pair of hadronic Zero Degree Calorimeters (ZDCs) located $\pm18$~m 
from the detector center and at zero degrees relative to the beam axis.

STAR employs several subsystems for particle tracking and
calorimetry \cite{NIM-RHIC}. The central Time Projection Chamber (TPC) tracks charged 
particles with $\sim$85$\%$ efficiency
over the full azimuth, $ 0 < \phi < 2 \pi $, and pseudorapidity range, $| \eta | < $1.0, 
and falling to $\sim$50$\%$ efficiency
at $| \eta | \sim$ 1.3. Surrounding the TPC is a Barrel Electromagnetic Calorimeter (BEMC) 
which consists of
4800 Pb-scintillator towers covering the full azimuth over the range $-0.98 < \eta < +0.98$ 
in its final configuration.
Each tower subtends an area of solid angle ($\Delta\phi \times \Delta\eta$) = ($0.05 \times 0.05$).
The positive $\eta$ side of STAR (west end) is covered by an additional 720 Pb-scintillator
towers comprising the Endcap Electromagnetic Calorimeter (EEMC), which extends
calorimeter coverage over the full azimuth for the pseudorapidity range
$1.08 < \eta < 2.0$. Both electromagnetic calorimeters are $\sim$20 radiations lengths and $\sim$1 
strong interaction length deep. Fast signals from the calorimeter
towers are processed to classify triggers for events of interest.
The reader is referred to Ref.~\cite{NIM-RHIC} for a comprehensive description of the STAR detector.

The measurements presented here were taken over two different running periods during the years 
2005 and 2006. 
In 2005, only the west half of the BEMC ($0<\eta<0.98$) was available.
Between the data taking in 2005 and 2006, the BEMC commissioning was completed.
This provided a more complete and robust picture of jets in our detector by doubling the acceptance and 
enabling measurements with jet cone radii, $R=\sqrt{\Delta\phi^2~+~\Delta\eta^2} = 0.7$,
which is larger than the value $R=0.4$ that was used in our earlier analyses~\cite{Abelev:2006uq,Abelev:2007vt}.

\subsection{Triggers and Data Sets}
A minimum bias (MB) trigger was defined to be a coincidence between any pair of BBC tiles from opposite 
sides of the collision region. This trigger has been shown to accept about 87$\%$ of the 
non-singly diffractive $pp$ 
cross section \cite{BBCref}. A redundant set of scalers recorded BBC tile and plane hits for 
each RHIC beam crossing, 
allowing the BBCs to be used as local luminosity monitors and polarimeters. The scaler system 
also recorded numerous 
combinations of hit conditions, including hits in the BBC on opposite sides of the interaction 
region in 15 unequal intervals 
of the time difference between the hits. The intervals were chosen so as to give fine granularity 
for beam-beam collisions 
that occurred near the center of the detector ($z$=0) and coarser granularity for events away 
from the center. 
We analyzed events from the intervals that correspond to a collision vertex selection along the beam direction of 
approximately $\pm60$~cm from the center of the detector. In this way, we matched the 
conditions for event 
selection with the conditions used in determining the relative luminosity for different 
spin combinations. 
The MB trigger was heavily prescaled to contribute only a few percent of the recorded data.

\begin{figure}
\includegraphics[width = 0.45\textwidth]{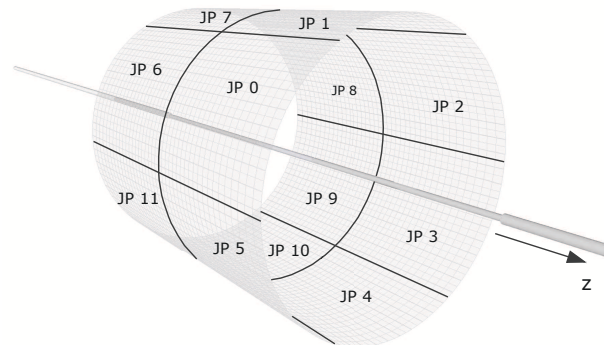}
\caption{\label{jetpatch} Schematic diagram showing the location of the fixed jet patches and
calorimeter towers in the STAR Barrel Electromagnetic Calorimeter.}
\end{figure}

Triggers for the selection of events with jets were constructed by requiring substantial energy
to be present in the BEMC. A High Tower (HT)
trigger was defined by requiring a BBC coincidence plus at least one
BEMC tower with a transverse energy greater than a given threshold. A
Jet Patch (JP) trigger required a BBC coincidence, plus a transverse electromagnetic energy in a
region of $\Delta\eta~\times~\Delta\phi~=~1.0~\times~1.0$ exceed a given
threshold. The locations of the jet patches were fixed
by hardware, with 12 such patches in the barrel calorimeter as shown in Fig.~\ref{jetpatch}.

In 2005, data were taken with a mixture of HT and JP triggers with
different thresholds. The low(high) HT1(2) triggers required
each accepted event to have at least one BEMC tower with transverse
energy $E_T > 2.6(3.5)$~GeV. The low(high) JP1(2)
trigger thresholds were set to $E_T > 4.5(6.5)$~GeV. The HT1 and JP1 triggers were prescaled.
There was considerable overlap among the triggers, with approximately half of the jets
contained in the JP2 trigger sample.

In 2006, the JP trigger was operated with a threshold of $E_T > 7.8$~GeV early in the 
run, including the entire transverse polarization period. The threshold was then increased 
to 8.3 GeV for most of the longitudinal polarization period. Two additional triggers were 
also used. The first was a HT trigger that required $E_T > 5.4$~GeV. The second was a refinement 
of the HT trigger (HTTP) that required a
high tower to exceed a threshold of 3.8 GeV, with an additional
requirement of $E_T > 5.2$~GeV in the $3 \times 3$ array of towers centered on the high tower.

The integrated luminosity for longitudinally polarized beam was 2.1 pb$^{-1}$ in 2005 
and 5.5 pb$^{-1}$ in 2006. The integrated luminosity for transversely polarized beam was 1.8 pb$^{-1}$ during 2006.

\section{Jets}
Jets measure energy flow and are observable as composites of measured particle momenta.
In the discussion below, we distinguish three categories: \emph{parton}, \emph{particle}, and \emph{detector} jets.
Our jet-finding and reconstruction method uses the approach and algorithm adopted from Ref.~\cite{CDF_jets}.
\emph{Parton} jets are constructed from quarks and gluons either in theoretical calculations or in
Monte Carlo simulated events prior to hadronization.
\emph{Particle} jets are constructed from the momenta of stable particles in Monte Carlo simulated events after hadronization.
\emph{Detector} jets are constructed from real or simulated data.
An important difference compared to many other experiments is that the charged hadron momenta are measured in STAR
with the TPC, rather than by a hadron calorimeter.

We use comparisons of parton-to-particle and particle-to-detector jets to quantify the
corrections needed to account for contributions arising from spectator partons, effects
of the underlying event in $p+p$ collisions, undetected energy from, for example, neutrons, $K_L$ 
and neutrinos, particle scattering out of the jet
cone due to the hadronization process, bias and resolution effects introduced by our triggers and detectors, 
and uncertainties in the relative contributions of different partonic processes that
result from uncertainties in the parton distribution functions.

\subsection{Jet-Finding and Reconstruction}
Detector jets are reconstructed in this analysis using a midpoint
cone algorithm \cite{JET-BLAZEY} that combines charged tracks from the TPC
and tower hits from the electromagnetic calorimeters. A primary vertex position
is defined from the intersection of two or more charged tracks with the known transverse
position of the beams. To be included in jet reconstruction, tracks are
required to have a transverse momentum greater than 200~MeV/$c$, while tower hits
must have a transverse energy $E_T$ exceeding 200~MeV. Charged tracks are also required 
to contain $>$20 ($>$12) fit points in the TPC for 2005 (2006) and $>$51\% of the maximum 
number of fit points allowed by the TPC geometry and active electronic channels. They are 
also required to have a distance-of-closest-approach to the primary 
vertex of $dca <$ 3~cm. For the 2006 data, an additional $p_T$-dependent constraint was imposed on the 
transverse distance of the track from the beam line ($dcaD$): $dcaD <$ 2 cm 
for $p_T <$ 0.5 GeV/$c$, $dcaD <$ 1 cm for $p_T >$ 1 GeV/$c$, and an interpolated cut in between.
The tracking capabilities of the TPC generally allow a determination of the $dca$ value with a resolution of
0.2-0.3~cm. These cuts are imposed to reduce pileup and background tracks in the data.
Particles measured as TPC trajectories are assumed to be charged pions, whereas energy deposits
in the electromagnetic calorimeters are assumed to be photons.
To reduce double counting of energy, the average energy that a minimum-ionizing particle (MIP) would deposit
is subtracted from the calorimeter tower energy if a TPC track points to the tower. 
If the calculated MIP energy deposition
exceeds the energy observed in such a tower, the tower energy is discarded.

%The midpoint cone algorithm begins by defining
%hits greater than a threshold energy to be seeds, and a proto-jet to be
%the collection of four-momenta inside a cone of predefined radius, $R$, whose axis coincides with
%the $E_T$-weighted centroid of the four-momenta. The first step is to
%sort the list of seeds by $E_T$ in descending order and find proto-jets
%around the seeds. As four-momenta are associated with proto-jets, they are
%removed from the list, and subsequent proto-jets are composed of four-momenta
%that are not part of any proto-jet. In the second step, after reaching the end of the four-momentum list,
%the algorithm is repeated using the list of midpoints of all pairs of proto-jets.
%It uses the midpoints as seeds and finds further proto-jets around the midpoints.
%The midpoint proto-jets share energy with proto-jets found in the first step.  

The midpoint cone algorithm begins by collecting a list of tracks and 
tower hits with transverse momentum/energy greater than a set threshold 
value.  These tracks and towers serve as seeds for the initial jet cones 
or proto-jets, defined as the collection of track and tower four-momenta 
inside of a cone of radius $R$, whose axis coincides with the $E_T$-weighted 
centroid of the proto-jet four-momenta. Additional proto-jets are 
formed from the midpoints between seeds and added to the list.   At this 
point, a single track or tower may contribute to several proto-jets. 
 
Next, the algorithm decides whether to split or merge two proto-jets
%The third step is to split/merge the proto-jets. 
%The algorithm decides whether to split or merge two proto-jets 
that have common four-momenta based on the
fraction of energy shared by the two proto-jets. If the fraction is smaller
than a specific value (0.5), the proto-jets are split into two jets and the
shared four-momenta are assigned to the closest jets. If the fraction is
greater than 0.5, the proto-jets are merged into a single jet. For 2005 data, the jet cone
radius was chosen to be $R$=0.4.  The cone radius was increased to $R$=0.7 for 2006 data
to increase the efficiency and minimize the sensitivity to differences in the fragmentation of quark vs.\@ gluon jets.
Parameters used in the jet-finding algorithm are summarized
in Table~\ref{TABLEM}.

\begin{table}[ht]
\begin{center}
\caption{\label{TABLEM} Midpoint Cone Algorithm Parameters}
\begin{tabular}{c|c|c}
\hline
\hline
Parameter & 2005 Jet-finding & 2006 Jet-finding \\
\hline
Cone Radius (rad) & 0.4 & 0.7 \\
Seed $E_T$ Threshold & 0.5 GeV & 0.5 GeV \\
Assoc. $E_T$ Threshold & 0.1 GeV & 0.1 GeV \\
Split/Merge Fraction & 0.5 & 0.5 \\
Track $p_T$ Threshold & 0.2 GeV/$c$ & 0.2 GeV/$c$ \\
Tower $E_T$ Threshold & 0.2 GeV & 0.2 GeV \\
%Minimum Track Hits & 20 & 12 \\
%\hline
%Track $\eta$ & $|\eta| < 1.6$ & $|\eta| < 2.0$ \\
%\hline
Jet $\eta$ & $0.2 < \eta < 0.8$ & $-0.7 < \eta < 0.9$ \\
Jet $p_T$ & $> 5.0$ GeV/$c$ & $>5.0$ GeV/$c$ \\
\hline
\hline
\end{tabular}
\end{center}
\end{table}

\subsection{Event and Jet Cuts}
Events were removed from this analysis in the absence of a valid polarization measurement, relative luminosity
value or BBC vertex information, or if the event failed the off-line verification of the trigger requirements.  
Events were also eliminated if the bunch identification tagged them as
originating from non-colliding, diagnostic (or ``kicked'') bunches. Kicked bunches are special bunches whose betatron orbits are deliberately amplified in order to give a large amplitude signal to a beam position monitor.  Timing measurements
of the kicked bunches give more precise measurements of the energy of the beam.
%Events were subjected to cuts requiring polarization, bunch spin identification (to eliminate 
%non-colliding, diagnostic, and `kicked' bunches), relative luminosity, BBC vertex information, 
%and off-line verification of trigger requirements,
After these requirements, longitudinal data samples of 4.6M events were obtained in both 2005 and 2006.
  
Many of the events contain two or more reconstructed jets. For the spin asymmetry measurements, 
only those triggered jets that contain a trigger tower or point to a triggered jet patch are included. 
Events that contain both a triggered jet and a non-triggered jet were included in the sample used to estimate the 
beam-gas background fraction (see next section).
Further cuts, as described in the following sections, were imposed on the jets to eliminate backgrounds. 
After these requirements, the longitudinal data samples totaled 2.3(2.1)M jets in 2005(2006). 2\%(4\%) of the events 
contained two jets, both of which passed all cuts.

\subsection{Background Events}

\subsubsection{Beam-Gas Events}
Energetic particles from beam-gas scattering and other non-collision background sources can pass 
through the beam-line shielding and then shower in the electromagnetic calorimeters. Most of these 
events lack signals in at least one of the BBCs, so they fail the trigger requirement. Typically, 
%they also fail our event cuts because they have no primary vertex within $-60<z_{vertex}<60$ cm. 
they also fail our event cuts because they have no primary vertex within the active region of the TPC. 
However, such events can be mistaken for jets if they occur during beam crossings that also 
contain minimum-bias collisions.

It is extremely rare to have two energetic background particles enter STAR concurrently, so non-collision 
background events almost never contain two or more reconstructed jets. Also, background jets typically 
exhibit an abnormally large fraction of electromagnetic to total energy, referred to as the neutral 
energy fraction (NEF), due to the lack of charged particle tracks that point back to an allowed vertex.
We use these features to identify the non-collision background contribution to our jet sample.

\begin{figure}
\includegraphics[width = 0.45\textwidth]{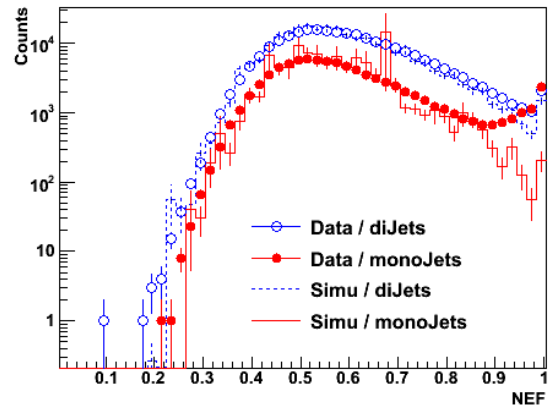}
\caption{\label{figure_1} (color online) Neutral energy fraction (NEF) distributions
for mono- and di-jets for 2006 data (symbols) and Monte Carlo simulation (histograms).
These events have uncorrected jet $p_T$ in the range $14.08~<~p_{T}~<~17.31$~GeV/$c$.}
\end{figure}

Figure \ref{figure_1} shows the neutral energy fraction distributions for our reconstructed 
jets when we divide them into two sub-samples. The di-jet events contain a second reconstructed 
jet with $\Delta\phi > \pi/2$. To enhance the di-jet statistics, the second jet is not required 
to satisfy the trigger independently. It is also allowed to fall within a larger pseudorapidity 
range than normal because it is not essential to reconstruct its energy precisely. The mono-jet 
events are the remainder. The measured neutral energy distribution for di-jet events is well 
described by PYTHIA events processed through a GEANT model of the STAR detector (see Sect. \ref{sect:simulations}). 
The mono-jet events are also well described except at large NEF, where a large enhancement due to 
non-collision background events is seen in the data.

\begin{figure}[hbt]
\includegraphics[width = 0.45\textwidth]{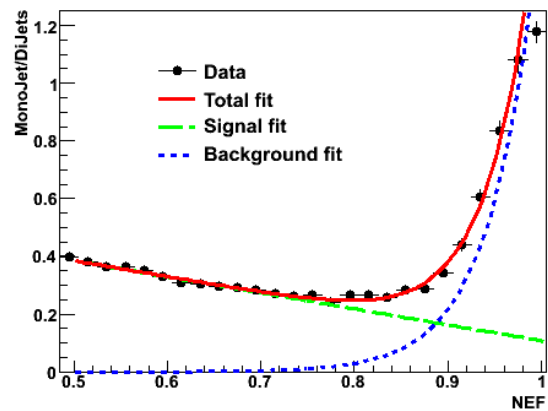}
\caption{\label{figure_2} (color online) Fits of the mono/di-jet ratio
vs.\@ NEF for 2006 data. The total fit (red solid curve) includes a decaying exponential for
the background (blue dotted curve) and a first order polynomial for the signal (green
dashed curve). These events have uncorrected jet $p_T$ in the range $14.08~<~p_{T}~<~17.31$~GeV/$c$.}
\end{figure}

The background component is determined from fits of the mono-jet to di-jet yield ratio as a 
function of NEF, as shown in Fig. \ref{figure_2}. Signal events have a slowly varying ratio, 
which is well described by a linear dependence in Monte Carlo simulations. The charged energy 
that is reconstructed in 
non-collision background events arises from tracks that accidentally point toward the calorimeter 
energy deposition.  These tracks are typically associated with the minimum-bias collision 
that was necessary to satisfy the trigger and produce a primary vertex. We find that this component 
can be fit with an exponential function.

The background mono-jets with NEF $>$ 0.95 were used to set an upper limit on the asymmetry in the 
non-collision backgrounds. We then established the cut for signal events to minimize the quadrature 
sum of the statistical and non-collision background uncertainties. For 2006, we required NEF $<$ 0.92. 
Less shielding was present in 2005, so we required NEF $<$ 0.90 for all bins except the lowest $p_T$ bin. 
It was difficult to isolate the background component in the fit for the lowest 2005 $p_T$ bin, leading 
to a very large uncertainty on any possible background asymmetry. Therefore, to be conservative we 
required NEF $<$ 0.85 for this bin.

\subsubsection{Electron-like Events}

%Jets sometimes contained a high energy electron (positron) coming from conversion of photons in the
%beam pipe, vertex detectors or support structures.  Due to the large uncertainties in
%simulating the distribution of this material and the energy sharing of the electron-positron pair,
%these jets were identified and eliminated from the analysis. 

%The energy from electrons (positrons) is double counted in jet reconstruction because, unlike charged hadrons, 
%the full $e^-(e^+)$ energy is reconstructed in the calorimeters. The jet finder will categorize a single
%energetic electron or positron as two high energy particles with similar four vector $E_T$ values from a TPC track 
%and EMC tower. High $p_T$ electron tracks in jets are readily identified by requiring the highest $p_T$ 
%track in the jet to project, within $\Delta\eta < \pm$0.03 and $\Delta\phi<\pm0.027$, to the location of 
%the highest $E_T$ tower. The ratio of the tower energy to track momentum was required to be less than 1.2. 
%Jets with $\sim$\,50\% of their total $E_T$ sum originating from towers associated with these tracks 
%were referred to as `electron-like' jets, 
%and were removed from the data sample. Agreement with simulations showed that these conditions effectively 
%eliminated such jets, which were about 20$\%$ of the inclusive jet sample.
%Most `electron-like' jets were found in HT triggered events, especially at low jet $p_T$. 

Initial studies of the NEF distributions for the HT and HTTP triggered events found an enhancement
in the jet yield for NEF in the range 0.4 to 0.5 in both the data and Monte Carlo simulations.  The enhancement,
which was particularly prevalent at low jet $p_T$, appeared primarily in the triggered mono-jet samples.  The
efficiency for observing the second jet in a di-jet increased with jet $p_T$, which indicated that the
likely cause was events where the jet energy had been significantly overestimated.  Further study determined that
the enhancement arose from events where a conversion electron or positron that fired the HT or HTTP trigger also
had a track reconstructed by the TPC. The jet finder will double-count the energy from such an electron or positron
because it categorizes the TPC track and EMC tower as two high-energy particles with similar 4-vector $E_T$ values.

A set of cuts was implemented to minimize the reconstruction bias associated with the double counting.  Jets in HT or
HTTP triggered events were discarded if the highest $p_T$ track in the jet projected, within $|\Delta \eta| < 0.03$ and
$|\Delta \phi| < 0.027$, to the location of the highest $E_T$ tower.  The ratio of the tower energy to track
momentum was required to be less than 1.2. Figure \ref{figure_1} shows that these cuts effectively removed these 
``electron-like'' events.

\subsection{Simulations}
\label{sect:simulations}

\begin{figure}
\includegraphics[width = 0.45\textwidth]{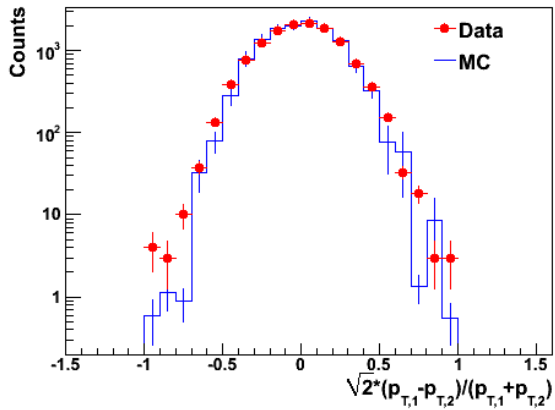}
\caption{\label{figure_4} (color online) Relative difference in $p_{T}$ for back-to-back di-jets for 2006 data. Both jets satisfy all cuts, 
including the trigger-matching requirement. These events have uncorrected jet $p_T$ 
in the range $14.08~<~(p_{T,1}+p_{T,2})/2~<~17.31$~GeV/$c$. For this case, we find $\sim$\,5\% discrepancy
between data and Monte Carlo simulations. Essentially all the bins match simulations to within 10$\%$.
The $p_{T}$ resolution is estimated to be $\sim$~23$\%$ from this graph.}
\end{figure}

\begin{figure}
\includegraphics[width = 0.45\textwidth]{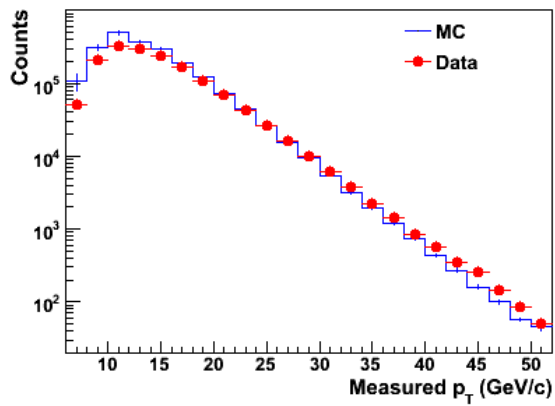}
\caption{\label{figure_3} (color online) Raw jet yield versus uncorrected jet transverse momentum
in 2006 data (points) compared with Monte Carlo simulations (histogram).}
\end{figure}

\begin{figure}
\includegraphics[width = 0.45\textwidth]{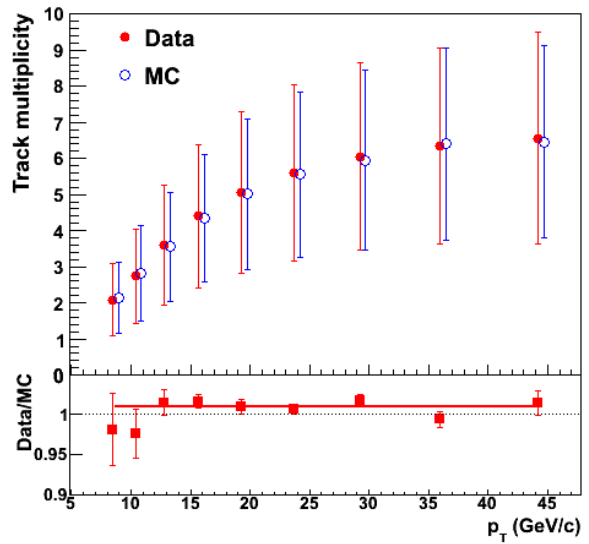}
\caption{\label{figure_6} (color online) Top Panel: Average track multiplicity in the reconstructed
jet $versus$ jet $p_T$ for 2006 data (solid symbols) and Monte Carlo simulations (open symbols). Vertical
bars represent the r.m.s.\@ width of the multiplicity distributions rather than the uncertainty
on the mean. Bottom panel: Ratio of data to Monte Carlo simulations. Vertical bars show the statistical uncertainties.}
\end{figure}

\begin{figure}[hbt]
\includegraphics[width = 0.45\textwidth]{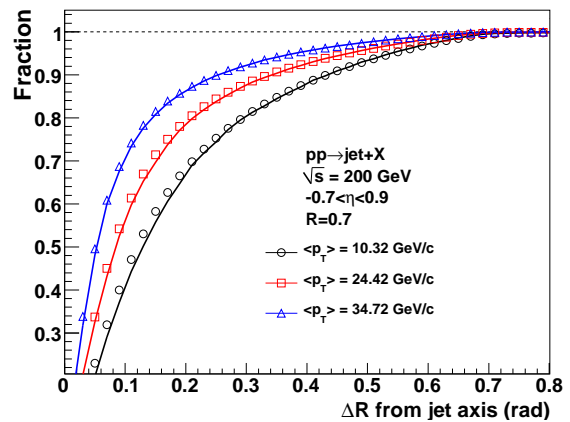}
\caption{\label{figure_8} (color online) Fraction of the total jet transverse energy found within a cone of radius $\Delta R$ 
centered on the reconstructed thrust axis, illustrating the jet profile. Data from 2006 (symbols) and
Monte Carlo simulations (curves) are shown for three different corrected jet $p_{T}$ bins.}
\end{figure}

\begin{figure}
\includegraphics[width = 0.45\textwidth]{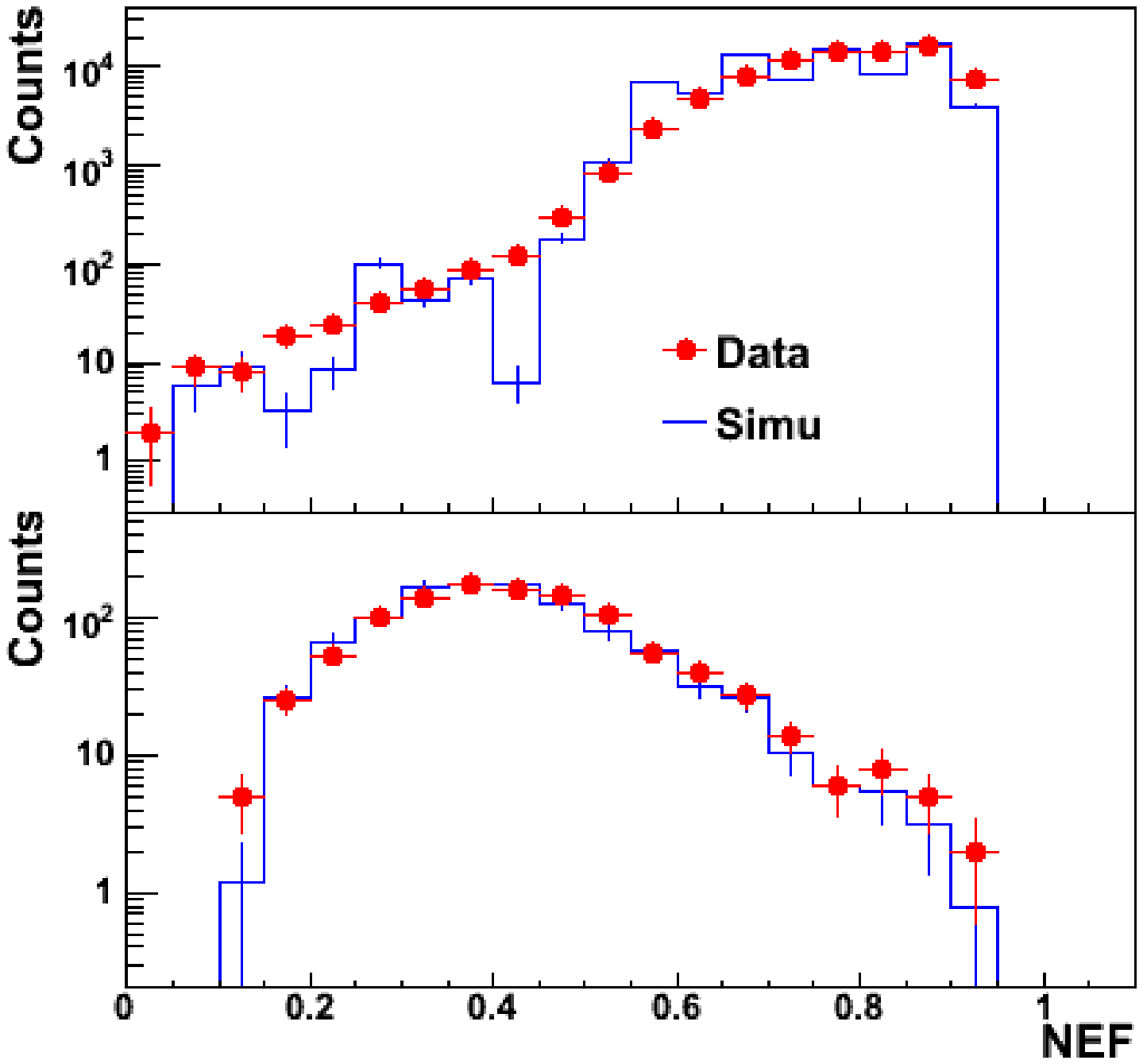}
\caption{\label{figure_5} (color online) Neutral energy fraction (NEF) for 2006 data (symbols) and Monte Carlo
simulations (histograms) for uncorrected jet $p_T$ in the ranges $7.6 < p_T < 9.3$~GeV/$c$ (upper panel) and $39.6 < p_T < 48.7$ GeV/$c$ (lower panel). These plots demonstrate the substantial bias introduced at low $p_T$ by triggering on only electromagnetic energy.}
\end{figure}

Monte Carlo events were generated using PYTHIA 6.205 \cite{Sjostrand:2000wi,Sjostrand:2001yu} (2005 data) 
and PYTHIA 6.410 \cite{Sjostrand:2000wi,Sjostrand:2006za} (2006 data)
with parameters adjusted to the CDF `Tune A'
settings \cite{Field:2005sa} and processed through the STAR detector
response package based on GEANT~3 \cite{geant}, including simulation of
the trigger electronics. The PYTHIA parameters corresponding to this
tune are listed in Table~\ref{TABLE0}. In order to achieve a satisfactory
simulation of the observed momentum balance for back-to-back jets (Fig.~\ref{figure_4}), we
incorporated an intrinsic parton transverse momentum of 1~GeV/$c$ into the default PYTHIA model. Small
discrepancies with the data for these model calculations may still be observed; however,
for the purpose of estimating the systematic uncertainties, the shapes of the data distributions
are reproduced sufficiently well by the simulations. We give further examples of the comparison of data
and Monte Carlo in Figs.~\ref{figure_3}-\ref{figure_8}
for the jet $p_T$ spectrum, track multiplicity and integrated transverse energy profile within the jets.
These figures use 2006 data and Monte Carlo
simulations. They require each jet to satisfy at least one of the HT, JP or HTTP trigger conditions, 
the same condition used in the asymmetry analysis.
The effect of the triggers on the character of the jets is seen most readily in
Fig.~\ref{figure_5}, which plots the NEF distribution of the jets for two different jet
momentum ranges. At low momenta, the calorimeter-based triggers preferentially select jets with
higher neutral energy fraction than jets at higher $p_T$. The systematic uncertainty associated with this
trigger bias will be discussed below.

\vspace{-0.3cm}
\begin{table}[!ht]
%\vspace{-0.1in}
\centering
\caption{\label{TABLE0} CDF Tune A parameters.}
\begin{tabular}{ c | c } % Column formatting, @{} suppresses leading/trailing space
\hline\hline
~Parameter~~ & Value \\
\hline
MSTP(51) & 7\\
MSTP(81) & 1\\
MSTP(82) & 4\\
PARP(82) & 2.0 \\
PARP(83) & 0.5 \\
PARP(84) & 0.4 \\
PARP(85) & 0.9 \\
PARP(86) & 0.95 \\
PARP(89) & 1800 \\
PARP(90) & 0.25 \\
PARP(91) & 1.0 \\
PARP(67) & 4.0 \\
\hline\hline
\end{tabular}
%\label{table:table0}
\end{table}

\vspace{-0.5cm}
\subsection{Jet Energy Scale}
Jet spin asymmetries are reported here as a function of the jet transverse momentum. 
However, a number of corrections must be made to the physical measurements to permit comparisons 
to the parton-level cross sections and jet $p_T$ definitions used in theoretical calculations. 
Some corrections are
best subsumed into the systematic uncertainties on the asymmetries themselves, while others are 
more naturally applied as shifts to the jet momenta.

\subsubsection{TPC and Calorimeter Calibration}
The TPC calibration proceeds in several steps~\cite{Gene}. First, the drift velocity, which 
is monitored during runtime via laser ionization of the gas, is determined to an accuracy of approximately 
0.03$\%$. Then, distortion of the tracks due to misalignments of the readout sectors, drift distortions in the 
magnetic field and construction imperfections are removed by examining extremely rigid tracks in different 
volumes of the TPC. Finally, the effects of space charge due to positive ion build-up at high event rates is 
monitored and corrected by examining the distance of closest approach to the presumed common vertex of an ensemble 
of tracks. This latter quantity is monitored as a function of instantaneous luminosity throughout the runs, and the 
corrections are updated periodically. The resulting hit errors on the tracks are in the range 300-550 $\mu$m. 
The overall momentum resolution of the TPC tracks is approximately 
$\Delta p_T /p_T \sim 0.01 ~+ ~0.005 p_T /$(GeV/$c$) for $p_T < 10$~GeV/$c$~\cite{Gene2}. The TPC tracking efficiency is $\sim85 \pm 5\%$.

The electromagnetic calorimeter was calibrated using a sample of identified electrons in the TPC data with momenta between 1.5 and 15 GeV/$c$ that satisfied strict geometrical and isolation conditions. 
The extrapolated trajectory was required to remain completely within a tower and the maximum energy of
the nearest neighbors was limited. 
%The entry point was required 
%to be within an angle of 0.004 radians from the center of the tower, and the maximum energy of the nearest 
%neighbors was limited to be less than half of the track energy. 
The tower calibration was then determined 
from the distribution of the ratio of energy observed in the towers divided by the momentum of the track. 
The variation of this ratio as a function of distance from the center of the tower was studied
in data and Monte Carlo, and a correction was applied to compensate for this
variation. The correction amounts to a maximum of 8\% at the edge of the
fiducial cut.

The total uncertainty in the calorimeter response to jets is estimated to be 4.8\% for the 
present analysis.  This includes the uncertainty from the electron calibration plus contributions 
from the uncertainty in the differential response of the calorimeter to hadronic vs.\@ electromagnetic 
energy and the ability of the Monte Carlo simulation to describe the precise light output from the 
scintillators when energy is deposited very close to the edge between two towers.  The nominal full 
scale gain for the individual calorimeter towers was changed from 28 GeV in 2005 to 60 GeV in 2006.  
Therefore, the calorimeter gain uncertainties are independent between the two years, even though the 
fractional uncertainties are equal.

\subsubsection{Jet $p_T$ Scale Corrections}
Our largest $p_T$ correction accounts for the difference in particle and detector
jet $p_T$ scales. The combination of a steeply-falling $p_T$ dependence in the
jet yield and a jet transverse momentum resolution of $\sim$\,23\% (seen in Fig.~\ref{figure_4}) 
causes substantial bin migration. 
Thus, on average, lower momentum jets are reconstructed as higher momentum jets. 
The most straight-forward method of dealing with this effect is to apply a $p_T$ shift to 
correct the average value of the detector jet $p_T$ within a bin.

This correction is calculated by comparing the $p_T$ centroid
for Monte Carlo simulations of particle versus detector jets, bin-by-bin and
for each trigger type, that are then combined to produce the $p_T$ shift for the data.
The main systematic uncertainties on this $p_T$ shift are due to
assumptions about the proportions of different partonic subprocesses contributing to the
jet spectrum.  The uncertainties were estimated by recalculating the
$p_T$ shifts for different subprocesses in PYTHIA, taking the maximum deviation for
any of them, then adding the statistical uncertainty in quadrature with this value.
The smaller cone size used in 2005, compared to 2006, results in a larger asymmetry in the
associated systematic error.
%The larger asymmetry of the errors in the 2005 data compared to 2006 data is due to the
%difference in cone sizes between the two years. 
%Figure~\ref{figure_7} gives this plot for the combination of all triggers for both
%2005 (solid points) and 2006 (open points). In this plot, the vertical bars represent
%the r.m.s.\@ width of the distributions within a bin rather than the uncertainty on the
%mean values. 
Table~\ref{TABLE1} (2005) and Table~\ref{TABLE2} (2006) give these corrections for both
years.  For each transverse momentum bin (first column) we give the mean detector jet $p_T$ (second
column) and the corrected mean particle jet $p_T$ (third column).
%The corrected mean $p_T$ for each bin (first column) is tabulated in the third column
%of Table~\ref{TABLE1} (2005) and Table~\ref{TABLE2} (2006) for both years. 
The fourth column
in each table gives the estimated uncertainties on the $p_T$ shift. The fifth column
arises from uncertainties in jet energy scale due to possible inaccuracies in
the calibration and performance of the TPC and EMC detectors. The comparison
of particle and detector jets was repeated for a variety of calibration ranges,
tracking inefficiencies, and detector states in order to estimate conservatively the range of
possible effects on the $p_T$ shift due to detector performance. 

\begin{table*}[ht]
\caption{\label{TABLE1} Jet transverse momentum bins and corrections for Y2005 data. For each
$p_T$ bin, the average detector jet $p_T$ and the corrected particle jet $p_T$ values are listed.
Also shown are the evaluated uncertainties on these $p_T$ values arising from the $p_T$-shift
procedure, from the residual uncertainties in our detector simulations, and from the
uncertainties on our hadronization and underlying event estimation and the unknown QCD scale. 
The sum in quadrature of these uncertainties is tabulated in the final column. All uncertainties are in units of GeV/$c$.}
%\centering
\begin{tabular}{c|c|c|c|c|c|c}
\hline\hline
Measured $p_T$ Range & $<p_T>$ & Corrected $<p_T>$ & $p_T$ Shift & Detector Sim. & Hadr/UE/QCD& Total \\
(GeV/$c$)& (GeV/$c$) & (GeV/$c$) &Uncertainty&Uncertainty&Scale Uncertainty & Uncertainty \\
\hline
5.00 - 6.15 & 5.58 & 5.32 & +0.18/-0.24 & $\pm$0.23 & +0.27/-0.15 & +0.40/-0.36 \\
6.15 - 7.56 & 6.86 & 6.30 & +0.11/-0.09 & $\pm$0.25 & +0.37/-0.28 & +0.46/-0.39 \\
7.56 - 9.30 & 8.43 & 7.06 & +0.36/-0.04 & $\pm$0.27 & +0.43/-0.34 & +0.62/-0.44 \\
9.30 - 11.44 & 10.37 & 8.67 & +0.20/-0.16 & $\pm$0.35 & +0.52/-0.39 & +0.66/-0.55 \\
11.44 - 14.08 & 12.76 & 10.73 & +0.09/-0.07 & $\pm$0.41 & +0.68/-0.54 & +0.80/-0.68 \\
14.08 - 17.31 & 15.70 & 13.08 & +0.08/-0.07 & $\pm$0.52 & +0.75/-0.54 & +0.92/-0.75 \\
17.31 - 21.30 & 19.31 & 16.00 & +0.19/-0.22 & $\pm$0.63 & +0.80/-0.50 & +1.04/-0.83 \\
21.30 - 26.19 & 23.75 & 19.39 & +0.30/-0.33 & $\pm$0.77 & +0.94/-0.53 & +1.25/-0.99 \\
26.19 - 32.22 & 29.21 & 23.57 & +0.38/-0.29 & $\pm$0.94 & +1.12/-0.61 & +1.51/-1.16 \\
32.22 - 39.63 & 35.92 & 28.07 & +0.58/-0.40 & $\pm$1.12 & +1.29/-0.65 & +1.80/-1.36 \\
\hline\hline
\end{tabular}
%\label{table:table1}
\end{table*}

\begin{table*}[ht]
\caption{\label{TABLE2} Jet transverse momentum bins and corrections for Y2006 data.
Details are the same as given in the caption for Table~\ref{TABLE1}.}
%\centering
\begin{tabular}{c|c|c|c|c|c|c}
\hline\hline
Measured $p_T$ Range & $<p_T>$ & Corrected $<p_T>$ & $p_T$ Shift & Detector Sim. & Hadr/UE/QCD& Total \\
(GeV/$c$)& (GeV/$c$) & (GeV/$c$) &Uncertainty&Uncertainty&Scale Uncertainty & Uncertainty \\
\hline
7.56 - 9.30 & 8.43 & 8.51 & +0.52/-0.37 & $\pm$0.35 & $\pm$0.51 & +0.81/-0.72 \\
9.30 - 11.44 & 10.37 & 10.32 & +0.30/-0.35 & $\pm$0.40 & $\pm$0.57 & +0.76/-0.78 \\
11.44 - 14.08 & 12.76 & 12.17 & +0.25/-0.23 & $\pm$0.46 & $\pm$0.76 & +0.92/-0.92 \\
14.08 - 17.31 & 15.70 & 14.41 & +0.08/-0.08 & $\pm$0.55 & $\pm$0.81 & +0.98/-0.98 \\
17.31 - 21.30 & 19.31 & 17.15 & +0.21/-0.17 & $\pm$0.66 & $\pm$0.86 & +1.10/-1.10 \\
21.30 - 26.19 & 23.75 & 20.45 & +0.13/-0.16 & $\pm$0.80 & $\pm$0.98 & +1.27/-1.28 \\
26.19 - 32.22 & 29.21 & 24.42 & +0.10/-0.12 & $\pm$0.97 & $\pm$1.17 & +1.52/-1.52 \\
32.22 - 39.63 & 35.92 & 29.41 & +0.22/-0.27 & $\pm$1.17 & $\pm$1.37 & +1.82/-1.82 \\
39.63 - 48.74 & 44.19 & 34.72 & +0.90/-1.22 & $\pm$1.38 & $\pm$2.03 & +2.61/-2.74 \\
\hline\hline
\end{tabular}
%\label{table:table2}
\end{table*} 

\subsubsection{Pile-up Corrections}
Pile-up refers to the rate-dependent correction for charged tracks and calorimeter hits that 
were accidentally added to a jet during reconstruction. The largest pile-up contribution came 
from out-of-time tracks that were nonetheless reconstructed within the 40~$\mu$s TPC read-out 
period. Additional sources included multiple events within the same bunch crossing or beam halo 
background that was coincident with a hard collision.

To estimate the size of this correction, during normal data-taking a
small fraction of events were taken with a random trigger, that is, data
taken during nominal beam crossings, but with no detector requirements.
These events are expected to contain the effects of the pile-up energy
alone, including the correct averaging over the instantaneous luminosity during the data-taking.

Jets were reconstructed in a sample of normal events.  The tracks and calorimeter hits from random
events in the same run were then added to these same normal events and a new set of jets were reconstructed.  
Separate average shifts were calculated for tracks and calorimeter hits and these average shifts 
were applied to the final jet $p_T$ spectra.
%Jets were first reconstructed in a
%sample of normal events. Zero-bias events from the same run were then examined to determine the additional 
%contribution pile-up would make to the reconstructed jet transverse momentum. 
For 2005(2006) this amounted
to a shift of 0.008(0.050) GeV/$c$ per jet. The larger correction for 2006 reflects the larger cone size used 
in the jet reconstruction and the higher instantaneous luminosity that was available.

\subsubsection{Hadronization and Underlying Event Corrections}
The $p_T$ shift applied above is still not complete.
There is a further difference in scale between
parton jet momentum and particle jet momentum that
may be divided into two partially compensating effects: underlying event (UE) and
out-of-cone (OOC)\cite{Bhatti:2005ai} fragmentation.

The OOC effect causes a reduction of the measured $p_{T}$ in particle
jets due to fragmentation and hadronization of the parent parton outside
of the jet cone. This correction may be sub-process dependent as quarks
are expected to have a harder fragmentation spectrum than gluons. The
underlying event causes an increase in the measured $p_{T}$ in particle
jets due to the inclusion of particles arising from interactions between
spectator partons in the proton remnants and from additional hard partonic
scatterings in the event. The UE is expected to be isotropic in
$\eta\times\phi$ space and has been found to be largely independent of
jet $p_{T}$.

The combined effect from the OOC and UE on the jet $p_{T}$ scale was
estimated by comparing reconstructed jets at the fragmented parton (FP)
and particle (PART) level in events generated by the PYTHIA 6.3 Monte Carlo
package \cite{Sjostrand:2000wi, Sjostrand:2003wg} with parameters set
to the CDF Tune A values. Jets at the FP stage contain only the fragmented
partons resulting from the scattered partons and the initial and final
radiation (MSTP(61), MSTP(71)). At the FP stage the underlying event and
hadronization (MSTP(81), MSTP(111)) are turned off. Note that MSTP(81) only
controls the multiple parton interaction component of the UE and does not
include effects from remnant interactions. Jets at the PART level contain
the stable, hadronized, final-state particles resulting from the interaction
in addition to any initial and final-state radiation. The reconstructed
jet $p_{T}$ scale at the PART level in simulations is comparable to the
experimentally measured jet scale after corrections for detector resolution
and trigger bias are included.

The total change in jet $p_T$ scale, $\Delta{p_T}=p^{FP}_{T}-p^{PART}_{T}$,
depends on the radius of the jet cone. 
%The smaller radius ($R$ = 0.4)
%resulted in a shift of $0.3 < \Delta{p_T} < 1.0$~GeV/$c$ for
%the range $7< p^{PART}_T < 22$~GeV/$c$, while the larger radius ($R$ = 0.7)
%produced a shift of $-1.2 < \Delta{p_T} < 0.3$~GeV/$c$
%over the momentum range $10 < p^{PART}_T < 40$~GeV/$c$.
%The deviation of the points from the dashed line in Fig.~10 indicate the
%sign and magnitude of $\Delta{p_T}$. 
Generally the shift is smaller for
jets reconstructed with the larger cone radius in 2006.
The reduced shifts at larger $p_T$ for $R$=0.7 indicates that
OOC effects become less important as the cone radius increases.
The lower $p_T$ behavior is dominated by
UE effects. As expected, the $\Delta{p_T}$ was found to
be sub-process dependent and larger for gluon jets. Therefore these
effects have been included as a systematic uncertainty, instead of a correction,
on the measured detector+trigger corrected jet $p_T$.
These uncertainties are given in the sixth columns of Table~\ref{TABLE1} and Table~\ref{TABLE2}.

\section{Spin Asymmetry Analysis}

\subsection{The Spin Asymmetry $A_{LL}$}
Experimentally, the double longitudinal spin asymmetry defined in Equation 1 was evaluated according to:
\begin{equation}
A_{LL}=
\frac{
\sum \left(P_1 P_2\right) \left( N^{++} - r N^{+-} \right)
}{
\sum \left(P_1 P_2\right)^2 \left( N^{++} + r N^{+-} \right)}
,
\end{equation}
where $P_{1,2}$ denote the measured beam polarizations, $N^{++}$ and $N^{+-}$ denote the
inclusive jet yields for equal and opposite proton beam helicity configurations, respectively,
and $r$ is the ratio of measured luminosities for the two helicity configurations.
Each sum is performed over runs that
last from 10 to 30 minutes so that the measurements are sampled on time-scales faster than
typical variations in the beam polarizations and relative luminosities.

\subsubsection{Beam Polarization}
The beams are injected and circulated as bunches in the RHIC rings with their spins oriented in the
vertical direction. Their polarizations are measured using Coulomb-nuclear interference polarimeters 
\cite{Jinnouchi:2004up,Okada:2005gu} that are calibrated against a polarized gas jet target \cite{Okada:2006dd} 
located at other interaction regions around the RHIC ring. The magnitude of the polarization is 
measured and monitored throughout the beam stores from these locations and is generally in the range 
from 50-55$\%$, with a statistical uncertainty of $\sim$\,1\%. 

%The fractional error quoted on the quantity $P_1 P_2$ is 9.4$\%$(8.3$\%$) for 2005(2006). 
%These errors represent an overall scale uncertainty on our measurements, and are common 
%with the polarization uncertainties of concurrent measurements performed by the PHENIX experiment.
%They are therefore quoted separately to facilitate comparison of different data sets and to
%identify correlated errors where possible.

\subsubsection{Relative Luminosity}
The relative luminosity for each polarization combination in a run was calculated from the sum of BBC
coincidences over a run, after sorting bunches for each spin combination.
Since these rates enter directly into the
expression for the asymmetries, care was taken to ensure these data
were consistent and systematically understood to a level commensurate
with the size of the asymmetry being measured.

The BBC and ZDC analog pulses are discriminated and the coincidence
signals are used as input signals to a time-to-amplitude converter, whose output is converted to a
4 bit time difference signal. For each beam crossing (every 106.5~ns),
4 bits for the BBC coincidence signal, 4 bits for the ZDC coincidence,
and 7 bits for the beam crossing number are distributed to a set of 4
redundant scaler boards. These scaler data are then examined for
statistical consistency; in general we find excellent agreement
among all methods of luminosity measurement. However, a small
fraction of runs ($<1\%$) were rejected due to inconsistencies among
the BBC measurements from different electronics channels.

As a further safeguard against detector
failure or subtle physical effects, the BBC relative luminosity measurements
were cross-checked against the ZDC measurements.
The BBCs are sensitive to the total non-singly diffractive $pp$
cross section by intercepting single charged particles over
a broad rapidity range at moderate transverse momenta \cite{BBCref}. This
hypothesis has been supported by PYTHIA calculations used in the
design of these detectors, and by direct measurements (Vernier scans),
which demonstrate that a very large fraction of the non-singly diffractive cross section is
indeed measured~\cite{drees}. The ZDCs detect
neutral particles like neutrons and $\pi^{0}$s close to the rapidity
of the beam, and are thus sensitive to types of
collisions which are very different from those sampled by the BBCs. The small acceptance of
the ZDCs limits the statistical precision of this comparison (for
the present data); however, this has the additional advantage of allowing
an examination of rate-dependence of the luminosity measurements as
well. The result of this comparison was consistent between the two
years and gives a conservative systematic uncertainty on the relative
luminosity of slightly less than 10$^{-3}$.

Corrections to the luminosity are expected
due to accidental coincidences and under-counting of multiple
interactions in a single beam crossing, as explained in
Ref.~\cite{CDF_LUM}. These effects have been examined for the
$relative$ luminosities encountered in the 2005/2006 runs and
found to be negligible compared to the systematic uncertainty
assigned to the relative luminosity by comparing BBC and ZDC measurements.

\subsection{Transverse Spin Asymmetries}
\label{sec:trans_spin}

In the 2006 run, STAR recorded 1.8~pb$^{-1}$ of jet data from
transversely polarized proton-proton collisions. These data
have been analyzed in order to measure the transverse spin asymmetries $A_{N}$,
$A_{\Sigma}$, and $A_{TT}$.

\begin{figure}
\includegraphics[width = 0.45\textwidth]{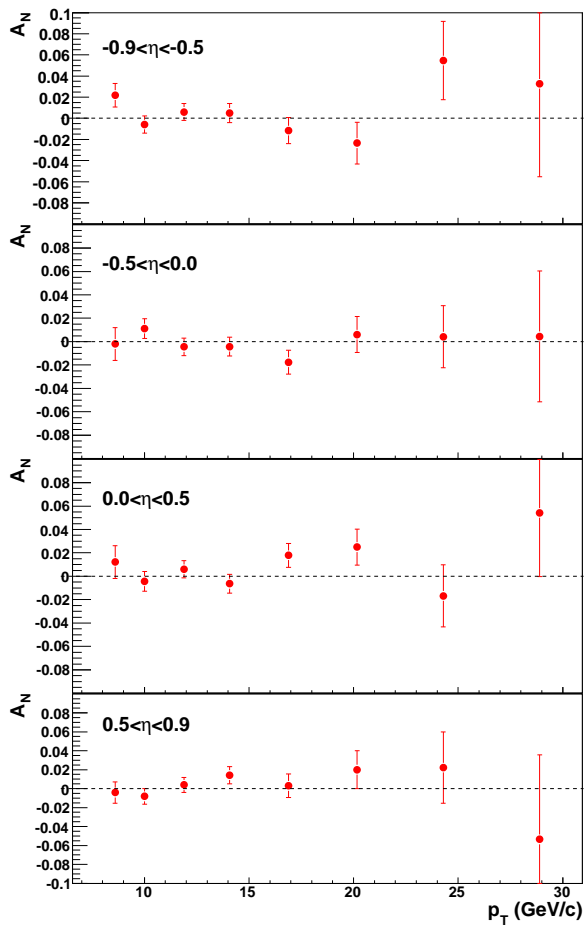}
\caption{\label{figure_10} (color online) $A_N$ as a function of the corrected mean $p_T$ for 2006
transverse data. The panels present $A_N$ for four different $\eta$ bins. $A_N$
is the left-right single-spin asymmetry for a transversely polarized beam. The errors shown combine the 
statistical uncertainties, which dominate, with all systematic uncertainties except trigger and reconstruction bias. 
See Sect. \ref{sect:A_N_Results} for a discussion of the latter.}
\end{figure}

In a coordinate system where the positive $z$ axis and pseudorapidity are
defined by the momentum direction of beam~1, the polarization directions of the beams are along
the $y$ axis, and the azimuthal angle $\phi$ is defined relative to the $x$ axis, we
can write the jet production cross section for the two transversely
polarized protons as in Ref.~\cite{meyer}:
\begin{equation}
\begin{aligned}
d\sigma_{pol} & / d\sigma_{unpol} = 1 + P_1 P_2 \cdot A_{\Sigma }(\eta, p_{T})\\
&+ \cos(\phi )\cdot [ P_1 \cdot A_{N}(\eta, p_{T} ) - P_2 \cdot A_{N}(-\eta, p_T ) ]\\
&+ ~~~ P_1 P_2 \cdot \cos(2 \phi ) \cdot A_{TT}(\eta,p_T )\\
\end{aligned}
\end{equation} 
Additional transverse spin asymmetries can be defined related to particle correlations within a jet \cite{Yuan08,DAlesio11} that are beyond the scope of this paper.

The extraction of the asymmetries proceeded as follows. The single-spin asymmetry, $A_N$, was 
determined by combining the spin directions for one beam to approximate an unpolarized `target'. 
The single-spin asymmetry for each beam was determined separately, using the cross-ratio technique \cite{Ohlsen73}, 
and the results combined. The statistical precision for this measurement was sufficient to allow measurements 
as a function of jet transverse momentum in 4 bins of pseudorapidity relative to the polarized 
beam, as shown in Fig.~\ref{figure_10}.

\begin{figure}
\includegraphics[width = 0.45\textwidth]{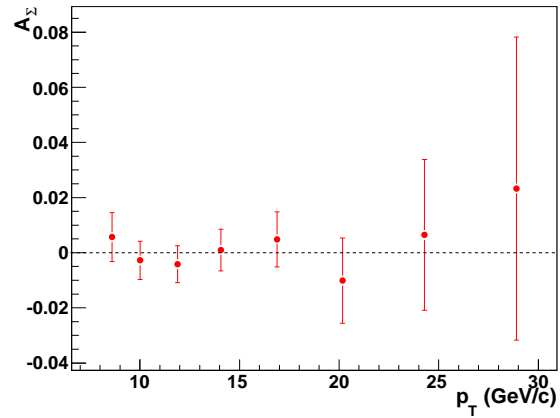}
\caption{\label{figure_11} (color online) $A_{\Sigma}$ versus corrected mean $p_T$ for 2006 transverse data, averaged over the
range $| \eta | <$~0.8.  $A_{\Sigma}$,
the transverse double-spin asymmetry for transversely polarized beams, is defined in more
detail in the text.}
\end{figure}

\begin{figure}
\includegraphics[width = 0.45\textwidth]{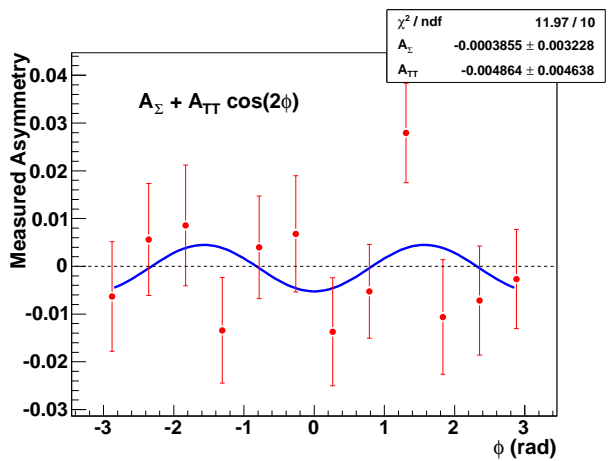}
\caption{\label{figure_12} (color online) Measured transverse double-spin asymmetry versus azimuthal angle $\phi$
for 2006 transverse data. The points are fit with the function $A_{\Sigma} + A_{TT}\cos(2 \phi )$.  
These data are averaged over 7.5~$< p_T < $40 GeV/$c$ and $| \eta | <$~0.8.}
\end{figure}

The double-spin asymmetry $A_{\Sigma}$ was determined by averaging over 
the entire range of pseudorapidity and azimuth, using analysis procedures identical to those for $A_{LL}$. 
The results for $A_{\Sigma}$ are given in Fig.~\ref{figure_11}. We are not aware of any theoretical predictions for $A_{\Sigma}$.  However, it plays an important role in the estimation of the systematic uncertainty in $A_{LL}$ due to residual transverse spin components in the beam (see Sect.\@ \ref{sec:resid_trans}).
A statistically significant 
measurement of $A_{TT}$ could be made only by averaging over the full data set and extracting the coefficient 
of the $\cos(2 \phi )$ dependence on the azimuthal angle from the fit shown in Fig.~\ref{figure_12}. 
We find $A_{TT} = -0.0049 \pm 0.0046$.  This precision is not yet sufficient to confront predictions of $A_{TT}$ due to quark transversity \cite{Soffer02}.

\subsection{Systematic Uncertainties}  

\subsubsection{Trigger and Reconstruction Bias}
Jet $p_{T}$ resolution effects shown in Fig.~\ref{figure_4} cause averaged
shifts in the jet $p_{T}$ scale as discussed in the previous section.
Additionally, jet events are selected based on neutral energy triggers,
that preferentially select jets with characteristics that differ
from those of the unbiased jet distribution. For
example, for the same jet momentum, the HT trigger will preferentially fire on jets with
a high-energy leading particle while the JP trigger will fire on jets with
larger radii. The relative proportions of quark-quark, quark-gluon and
gluon-gluon interactions in an unbiased sample of events are fixed 
predominantly by the (well-determined) unpolarized
structure functions. However, effects of the trigger and jet reconstruction can 
change the relative proportions 
in our measured sample, and this can bias our measurements of $A_{LL}$ and $A_N$.

Calculation of the biases introduced by our trigger and jet reconstruction
further depends on assumptions of the polarized parton structure
functions. Our calculations must account for the
uncertainty in polarized gluon contributions and, to a lesser degree, the
associated uncertainties in polarized quark and sea contributions.

Parameterizations of the polarized parton distribution functions are combined with
PYTHIA parton kinematic variables to generate predictions of $A_{LL}$ vs.\@
$p_{T}$ specific to a particular model at both the particle and
detector levels. A broad range of polarized parton distribution functions were adopted 
for these calculations. Eventually, only those that predict distributions for $A_{LL}$ 
vs.\@ $p_T$ consistent with the general trends of our measured results were included in the bias estimate.
 
\begin{figure}
\includegraphics[width = 0.5\textwidth]{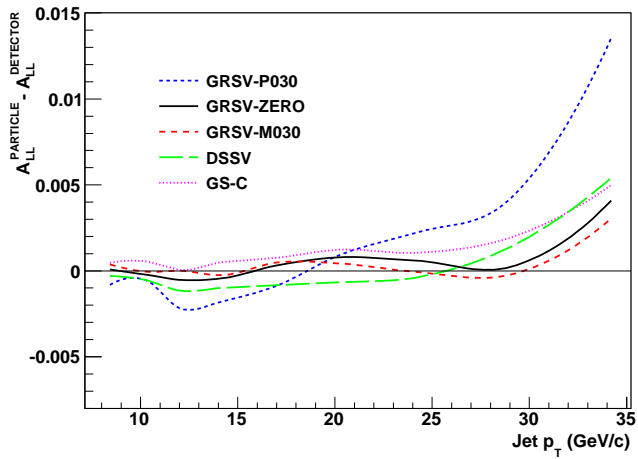}

\caption{\label{figure_9} (color online) Trigger and reconstruction bias estimates for the longitudinal
double-spin asymmetry $A_{LL}$ versus jet $p_T$ for a representative set of polarized parton 
distribution models \cite{gs},\cite{Gluck:2000dy}-\cite{dssv}.}
\end{figure} 

The method of calculating the trigger and reconstruction bias is then as
follows. Relative fractions of jets resulting from HT, JP, and
HTTP triggered events have been measured in data and found to agree
with simulations. The logical OR of the three trigger types is therefore used
to measure the trigger bias. For each $p_{T}$ bin and polarized parton distribution model, $A_{LL}$ is
calculated both at the particle and detector levels.
Detector $A_{LL}$ points are shifted as discussed above to
correct to the particle jet $p_{T}$ scale. The residual difference
between the particle jet $A_{LL}$ and the shifted detector jet $A_{LL}$
represents the bias for that model and $p_{T}$ bin. To
be conservative, the systematic uncertainty for each bin is assigned to be
the largest positive and negative difference of all the allowed
models. In Fig.~\ref{figure_9} we show the result of this calculation for five representative models.
The GRSV +0.3 and GRSV -0.3 models \cite{Gluck:2000dy, SVpriv} use the GRSV functional form for the gluon polarization, with the
integral fixed at the two respective values. DSSV \cite{dssv} is a recent fit that includes input
data from RHIC, in addition to DIS and semi-inclusive DIS. GS-C \cite{gs} is an early model that has a
large gluon polarization at low $x$. These five models 
span a range in gluon polarization that is wider than permitted by our results.
The systematic uncertainties as a function of $p_T$ are listed in the second column of
Tables~\ref{TABLE3} and \ref{TABLE4}.

\begin{table*}[ht]
\caption{\label{TABLE3} $p_T$-dependent systematic uncertainties for Y2005 data. The trigger and jet
reconstruction bias, non-longitudinal beam polarization, and beam-gas background
systematic uncertainties on the measured 2005 $A_{LL}$ are given. }
\vspace{0.1in}
\begin{tabular}{c|c|c|c}
\hline\hline
$p_T$ & Trigger bias and Jet Recon. & Non-longitudinal Pol. & Beam Gas Background \\
(GeV/$c$) & ($\times$ 10$^{-3}$) & ($\times$ 10$^{-4}$) & ($\times$ 10$^{-4}$) \\
\hline
%5.60 & -1.67/+1.67 & $\pm$4.97 & $\pm$13.50 \\
%6.14 & -1.47/+1.29 & $\pm$2.49 & $\pm$8.07 \\
%6.83 & -1.44/+1.10 & $\pm$1.54 & $\pm$7.97 \\
%8.67 & -1.80/+2.66 & $\pm$1.20 & $\pm$8.40 \\
%10.34 & -1.24/+2.32 & $\pm$1.16 & $\pm$6.82 \\
%12.89 & -1.34/+2.50 & $\pm$1.31 & $\pm$5.34 \\
%15.65 & -1.88/+2.86 & $\pm$1.74 & $\pm$4.55 \\
%19.30 & -2.74/+2.74 & $\pm$2.69 & $\pm$3.51 \\
%23.48 & -3.91/+3.91 & $\pm$4.75 & $\pm$0.50 \\
%27.94 & -3.44/+5.60 & $\pm$9.56 & $\pm$0.00 \\
5.32 & -1.67/+1.67 & $\pm$4.97 & $\pm$13.50 \\
6.30 & -1.47/+1.29 & $\pm$2.49 & $\pm$8.07 \\
7.06 & -1.44/+1.10 & $\pm$1.54 & $\pm$7.97 \\
8.67 & -1.80/+2.66 & $\pm$1.20 & $\pm$8.40 \\
10.73 & -1.24/+2.32 & $\pm$1.16 & $\pm$6.82 \\
13.08 & -1.34/+2.50 & $\pm$1.31 & $\pm$5.34 \\
16.00 & -1.88/+2.86 & $\pm$1.74 & $\pm$4.55 \\
19.39 & -2.74/+2.74 & $\pm$2.69 & $\pm$3.51 \\
23.57 & -3.91/+3.91 & $\pm$4.75 & $\pm$0.50 \\
28.07 & -3.44/+5.60 & $\pm$9.56 & $\pm$0.00 \\
\hline\hline
\end{tabular}
%\label{table:table3}
\end{table*}

\begin{table*}[ht]
\caption{\label{TABLE4} $p_T$-dependent systematic uncertainties for Y2006 data.
Details are the same as given in the caption for Table~\ref{TABLE3}. }
\vspace{0.1in}
\begin{tabular}{c|c|c|c}
\hline\hline
$p_T$ & Trigger bias and Jet Recon. & Non-longitudinal Pol. & Beam Gas Background \\
(GeV/$c$) & ($\times$ 10$^{-3}$) & ($\times$ 10$^{-4}$) & ($\times$ 10$^{-4}$) \\
\hline
%8.09 & -2.00/+3.37 & $\pm$0.90 & $\pm$15.17 \\
%9.53 & -1.07/+1.97 & $\pm$0.71 & $\pm$7.66 \\
%11.50 & -1.26/+1.99 & $\pm$0.68 & $\pm$4.92 \\
%13.91 & -0.58/+1.11 & $\pm$0.77 & $\pm$3.43 \\
%16.78 & -0.43/+0.70 & $\pm$1.02 & $\pm$3.57 \\
%20.25 & -0.72/+1.52 & $\pm$1.58 & $\pm$4.48 \\
%24.49 & -1.03/+3.92 & $\pm$2.79 & $\pm$7.52 \\
%29.70 & -1.57/+5.46 & $\pm$5.61 & $\pm$7.92 \\
%34.96 & -2.88/+6.93 & $\pm$12.85 & $\pm$5.01 \\
8.51 & -2.00/+3.37 & $\pm$0.90 & $\pm$15.17 \\
10.32 & -1.07/+1.97 & $\pm$0.71 & $\pm$7.66 \\
12.17 & -1.26/+1.99 & $\pm$0.68 & $\pm$4.92 \\
14.41 & -0.58/+1.11 & $\pm$0.77 & $\pm$3.43 \\
17.15 & -0.43/+0.70 & $\pm$1.02 & $\pm$3.57 \\
20.45 & -0.72/+1.52 & $\pm$1.58 & $\pm$4.48 \\
24.42 & -1.03/+3.92 & $\pm$2.79 & $\pm$7.52 \\
29.41 & -1.57/+5.46 & $\pm$5.61 & $\pm$7.92 \\
34.72 & -2.88/+6.93 & $\pm$12.85 & $\pm$5.01 \\
\hline\hline
\end{tabular}
%\label{table:table4}
\end{table*}

\subsubsection{False Asymmetries from Residual Transverse Spin Effects}
\label{sec:resid_trans}
False asymmetries that mimic our $A_{LL}$ signal
can arise from a combination of physical and experimental sources.
To obtain longitudinal collisions at STAR, the transversely polarized beams
are rotated to the longitudinal direction, then back again to transverse on
either side of the interaction region by a pair of helical dipoles
known as spin rotators. Inaccuracies in the adjustment
of the spin rotator currents leave small transverse components
for both beams in the collision region. Our transverse spin asymmetry measurements 
allow us to put stringent limits on the associated false asymmetries.

The transverse asymmetries in the central rapidity region are expected (and measured) 
to be small. However, 
in the presence of non-longitudinal polarization components, the asymmetry $A_{\Sigma}$ 
can contribute directly to the observed $A_{LL}$ signal.
Local measurements of the transverse polarization components of both beams during
$longitudinal$ running were made by examining the single-spin
asymmetries observed in BBC tile hits. The transverse single-spin asymmetry ($A_N$)
has been reported previously for this detector \cite{BBCref,kiryluk:2003aw}.
As given in these references, it is in
the range of $\sim$~6-7 $\times 10^{-3}$ and can be
calibrated to a high accuracy during $transverse$ running. Because
the BBC is a highly segmented detector, combinations of up/down and
left/right scatterings can be used to measure the transverse polarization components
for both beams. These measurements were made continuously
through the data-taking. The residual transverse components for both beams during the 
nominally longitudinal run
were weighted by integrated luminosity for different periods of adjustment for the
spin rotators. Denoting the angle of the polarization with the longitudinal axis as $\theta$,
values of $\tan( \theta )$ between 0.02-0.18 were measured, with an average magnitude equal 
to $\sim$\,0.1 for both beams.

Because measurements of $A_{\Sigma}$ were consistent with zero, we do not
make a correction to $A_{LL}$ for this contribution, but instead assign
a systematic uncertainty on our $A_{LL}$ measurements. We combined
the measurements of the transverse polarization components with the $uncertainties$
on the measurements of $A_{\Sigma}$ in each momentum bin
to give a conservative (maximal) estimate of the systematic uncertainty:
\begin{eqnarray}
\delta A_{LL} & = & \left | \tan(\theta_{1})\tan(\theta_{2})\cos(\phi_{1} -
\phi_{2}) \times ~A_{\Sigma} \right | \nonumber \\
~ & \sim & |\theta_{1} \theta_{2} \delta A_{\Sigma}| .
\end{eqnarray}
where $\theta_1$($\theta_2$) and $\phi_1$($\phi_2$) are the polar and azimuthal angles of the
polarization vectors for beams 1 (2), respectively. In keeping with the spirit
of estimating this uncertainty conservatively, the value of $\cos(\phi_{1}-\phi_{2})$ was set equal to unity.
The uncertainty on $A_{LL}$ due to non-longitudinal components of the beams
as a function of $p_T$ is listed in the third column of Tables~\ref{TABLE3} and \ref{TABLE4}.

\subsubsection{Beam Gas Background}
%Minor systematic uncertainties in the measured asymmetries arise from the previously
%discussed effects. 
The systematic uncertainty on the residual beam background was
conservatively estimated to be the larger of either the measured effect or the
statistical uncertainty of the measured effect on the asymmetry. These numerical
values range from $\sim 15 \times 10^{-4}$ in the smallest $p_T$ bins to less than
half this value at higher $p_T$. This uncertainty as a function of $p_T$ is listed in the fourth column of
Tables~\ref{TABLE3} and \ref{TABLE4}.

\subsubsection{Polarization and Relative Luminosity}
Systematic errors arising from beam polarization and relative luminosity measurements are treated 
separately due to their correlated effects on the data. An error in the relative luminosity measurement 
%would result in an overall shift of the $A_{LL}$ data in the vertical direction, while an error in 
would result in a shift of the $A_{LL}$ data points by an additive constant, while an error in 
the measurement of the polarization magnitude would scale the magnitude of the $A_{LL}$ data. 
Therefore we quote these quantities separately.

The systematic uncertainty on the relative luminosity was determined by comparing the BBC and ZDC
measurements, which were found to be consistent at the $<10^{-3}$ level. The BBC-ZDC difference was used to
estimate the possible size of the uncertainty on $A_{LL}$ due to the errors on the relative
luminosity as $\delta A_{LL} \sim 9 \times 10^{-4}$ for both 2005 and 2006 data.

The fractional systematic uncertainty for the quantity $P_1 P_2$ quoted by the RHIC CNI polarimeter
group (common to all RHIC experiments) is $9.4\%$($8.3\%$) for 2005(2006)\cite{rhicspin}.
Polarization measurements from different years have contributions which may be identified
as either uncorrelated or correlated.  The total error was conservatively estimated by
assuming the latter portion to be 100$\%$ correlated from year to year.  The correlated
error in the normalization of beam polarizations comes mostly from an unpolarized molecular 
hydrogen background in the gas jet polarization measurement. 
These errors represent an overall scale uncertainty on our measurements, and are common 
with the polarization uncertainties of concurrent measurements performed by the PHENIX experiment.
They are therefore quoted separately to facilitate comparison of different data sets and to
identify correlated errors, where possible.

\subsection{False Asymmetries}
All measurements were examined as a function of time to ensure the absence of non-statistical variations.
In addition, the data for different beam and bunch combinations were combined to form parity-violating 
single and double spin asymmetries. These are expected to be highly suppressed and provide internally 
consistent cross-checks on the validity of the measurements.  Double spin asymmetries were formed from
the ``like-sign'' and ``unlike-sign'' bunch combinations, and single spin longitudinal asymmetries were 
formed for each beam direction. No false asymmetry was found to be significantly different from zero.

\section{Results and Discussion}

\subsection{$A_N$ Results}

\label{sect:A_N_Results}
In a recent model calculation, D'Alesio \textit{et al}.\@ \cite{DAlesio11} conclude 
that $A_N$ for inclusive jets at mid-rapidity arises solely from the gluon Sivers effect. 
Within their model, the leading contribution to the systematic uncertainty on our measured 
inclusive jet $A_N$ arises from the fact that our triggers have different efficiencies for 
detecting jets from either quark-quark, quark-gluon, or gluon-gluon scattering. 
%especially 
Although the latter two processes dominate at low jet $p_T$, the detector efficiency is
largest for quark-quark scattering.
%At low jet $p_T$, gluon processes dominate,the largest efficiency is for quark-quark scattering. 
This can lead to a measured $A_N$ that is smaller in magnitude than the true value. We have used our 
Monte Carlo simulations to estimate the size of this effect. We find that correcting for 
this bias would increase the magnitude of $A_N$ by up to 40\% for low-$p_T$ jets, 
dropping to 25\% at 15 GeV/$c$ and 15\% at 30 GeV/$c$.

D'Alesio \textit{et al}.\@ find that the current upper limit on the gluon 
Sivers distribution would lead to $|A_N|$ of 4-5$\%$ at $p_T$ = 8 GeV/$c$, 
dropping to $\sim$\,2.5\% at 15 GeV/$c$ \cite{DAlesio11}. The results in 
Fig.\@ \ref{figure_10} indicate that our measured $A_N$ is substantially smaller than these 
upper limits. Thus, they may provide new constraints on the magnitude of the 
gluon Sivers distribution in the proton.

\subsection{$A_{LL}$ Results}
\begin{figure}[htb!]
\includegraphics[width = 0.5\textwidth]{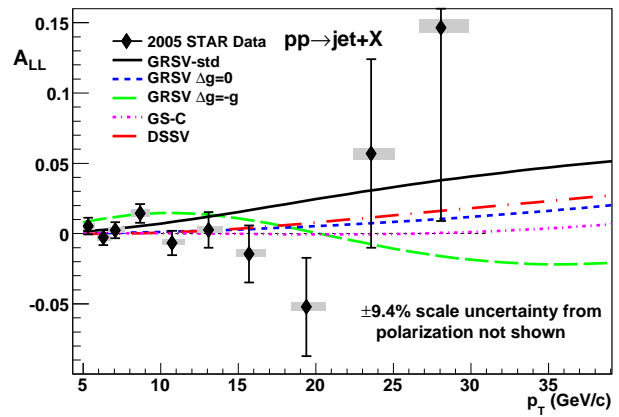}
\caption{\label{figure_13} (color online) $A_{LL}$ for inclusive jet production versus corrected jet $p_T$ for 2005 data.
For 2005 data we used a cone radius equal to 0.4 and a pseudorapidity range for the jet thrust axis of $0.2<\eta<0.8$.  The error bars are statistical.  The gray boxes show the systematic uncertainties.}
\end{figure}

\begin{figure}[t!]
\includegraphics[width = 0.5\textwidth]{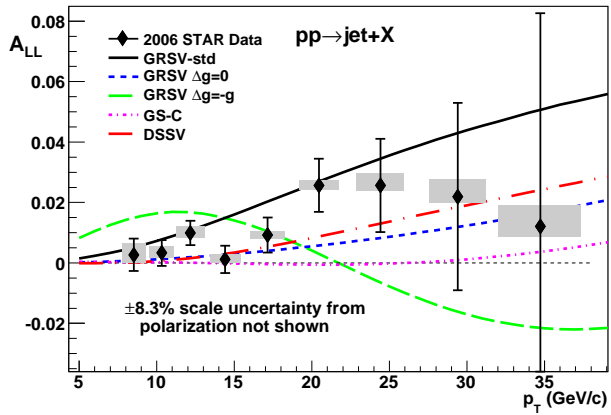}
\caption{\label{figure_14} (color online) $A_{LL}$ for inclusive jet production versus corrected jet $p_T$ for
2006 data. For 2006 data we used a jet cone radius equal to 0.7 and a 
pseudorapidity range of $-0.7<\eta<0.9$ to reflect the increased detector coverage.  The error bars are statistical.  The gray boxes show the systematic uncertainties.}
\end{figure}

The different detector geometries, triggers, jet definitions, and
measured pseudorapidity ranges of the
two different data taking years reported here have demanded independent evaluations of the
asymmetries and systematic uncertainties.
While the list of corrections from both years are the same,
individual items differ in magnitude and range from year to year. Furthermore, the different jet definitions and pseudorapidity intervals for the two years lead to different expectations for $A_{LL}$ from model calculations. We therefore
do not combine the results from different years, but present them
separately. Our results from year 2005 data are given in Fig.~\ref{figure_13} and Table~\ref{TABLE5}.
The corresponding results for year 2006 data are given in Fig.~\ref{figure_14} and Table~\ref{TABLE6}. Note the different scales on the vertical axes for the two figures.

%\vspace{-0.1in}

\begin{table}[ht!]
%\label{TABLE5}
\caption{\label{TABLE5} The final measured $A_{LL}$ and $p_T$ values from the 2005 data sample.
Data cover the range $0.2 < \eta < 0.8$ with a jet cone radius of $R$=0.4.
Statistical and systematic uncertainties are listed for $A_{LL}$ (note the systematic
uncertainty is asymmetric as described in the text). The $p_T$-dependent
corrections listed in Table~\ref{TABLE3} were combined in quadrature with the
systematic uncertainty of $9 \times 10^{-4}$ in relative luminosity to give the
total. The $p_T$ values shown are
the results after applying all $p_T$ corrections discussed in the text.}
\vspace{0.1in}
\begin{tabular}{c|c|c|c}
\hline\hline
$p_T$ & $A_{LL}$ & stat. err. & sys. err.\\
(GeV/$c$) & ($\times$ 10$^{-3}$) &($\times$ 10$^{-3}$) & ($\times$ 10$^{-3}$) \\
\hline
5.3 +0.4/-0.4& 5.3 & $\pm$5.9 & +2.4/-2.4\\
6.3 +0.5/-0.4& -2.7 & $\pm$5.4 & +1.8/-1.9\\
7.1 +0.6/-0.4& 2.4  & $\pm$5.7 & +1.7/-1.9\\
8.7 +0.7/-0.6& 14.3 & $\pm$6.7 & +2.9/-2.2\\
10.7 +0.8/-0.7& -6.7 & $\pm$8.7 & +2.6/-1.7 \\
13.1 +0.9/-0.8& 2.6 & $\pm$12.7 & +2.7/-1.7 \\
16.0 +1.0/-0.8& -14.6 & $\pm$20.3 & +3.0/-2.2 \\
19.4 +1.3/-1.0& -52.2 & $\pm$35.0 & +2.9/-2.9 \\
23.6 +1.5/-1.2& 56.9 & $\pm$67.1 & +4.1/-4.1\\
28.1 +1.8/-1.4& 146 & $\pm$138 & +5.7/-3.7\\

\hline\hline
\end{tabular}
%\label{table:table5}
\end{table}

\begin{table}[!]
\caption{\label{TABLE6} The final measured $A_{LL}$ and $p_T$ values from the 2006 data sample.
Data cover the range $-0.7 < \eta < 0.9$ with a jet cone radius of $R$=0.7.
Details are the same as given in the caption for Table~\ref{TABLE5}.
%Statistical and systematic uncertainties are listed for $A_{LL}$ (note the systematic
%uncertainty is asymmetric as described in the text). The $p_T$-dependent
%corrections listed in Table~\ref{TABLE4} were combined in quadrature with the
%systematic uncertainty of $9 \cdot 10^{-4}$ in relative luminosity to give the total. .The $p_T$ values shown are
%the results after applying all $p_T$ corrections discussed in the text.
}
\vspace{0.1in}
\begin{tabular}{c|c|c|c}
\hline\hline
$p_T$ & $A_{LL}$ & stat. err. & sys. err.\\
(GeV/$c$) & ($\times$ 10$^{-3}$) & ($\times$ 10$^{-3}$) & ($\times$ 10$^{-3}$) \\
\hline
8.5 +0.8/-0.7 & 2.7 & $\pm$5.3 & +3.8/-2.6\\
10.3 $\pm$0.8 & 3.3 & $\pm$4.3 & +2.3/-1.6\\
12.2 $\pm$0.9 & 9.9 & $\pm$4.1 & +2.3/-1.6\\
14.4 $\pm$1.0 & 1.2 & $\pm$4.5 & +1.5/-1.2\\
17.2 $\pm$1.1 & 9.2 & $\pm$5.8 & +1.2/-1.1\\
20.5 $\pm$1.3 & 25.7 & $\pm$8.8 & +1.8/-1.3\\
24.4 $\pm$1.5 & 25.6 & $\pm$15.4 & +4.1/-1.6\\ 
29.4 $\pm$1.8 & 22.0 & $\pm$31.0 & +5.6/-2.1\\
34.7 +2.6/-2.7 & 12.0 & $\pm$70.6 & +7.1/-3.3\\

%8.1 $\pm$0.7 & 2.7 $\pm$5.3 +3.8/-2.6\\
%9.5 $\pm$0.7 & 3.3 $\pm$4.3 +2.3/-1.6\\
%11.5 $\pm$0.9 & 9.9 $\pm$4.1 +2.3/-1.6\\
%13.9 $\pm$1.0 & 1.2 $\pm$4.5 +1.5/-1.2\\
%16.8 $\pm$1.1 & 9.2 $\pm$5.8 +1.2/-1.1\\
%20.3 $\pm$1.3 & 25.7 $\pm$8.8 +1.8/-1.3\\
%24.5 $\pm$1.5 & 25.6 $\pm$15.4 +4.1/-1.6\\ 
%29.7 $\pm$1.8 & 22.0 $\pm$31.0 +5.6/-2.1\\
%35.0 $\pm$2.5 & 12.0 $\pm$70.6 +7.1/-3.3\\

%8.09 $\pm$0.65 & 2.7 $\pm$5.3 +3.8/-2.6\\
%9.53 $\pm$0.72 & 3.3 $\pm$4.3 +2.3/-1.6\\
%11.50 $\pm$0.90 & 9.9 $\pm$4.1 +2.3/-1.6\\
%13.91 $\pm$0.99 & 1.2 $\pm$4.5 +1.5/-1.2\\
%16.78 $\pm$1.09 & 9.2 $\pm$5.8 +1.2/-1.1\\
%20.25 $\pm$1.27 & 25.7 $\pm$8.8 +1.8/-1.3\\
%24.49 $\pm$1.52 & 25.6 $\pm$15.4 +4.1/-1.6\\
%29.70 $\pm$1.81 & 22.0 $\pm$31.0 +5.6/-2.1\\
%34.96 $\pm$2.47 & 12.0 $\pm$70.6 +7.1/-3.3\\

\hline\hline
%\label{table:table6}
\end{tabular}
\end{table}

\subsection{Comparison to Theory}
The theoretical curves shown in Figs.\@ \ref{figure_13} and \ref{figure_14} are derived 
from NLO calculations of spin asymmetries based on the code of Jager et al.\,\cite{Jager:2004jh}. 
This code provides both the polarized and unpolarized proton-proton cross sections for an input 
cone of radius $R$ centered at rapidity $y$ and averaged over azimuth to $O(3)$ in $\alpha_{s}$. 
Expressions for all $ 2 \rightarrow 2$ and $2 \rightarrow 3$ processes were derived analytically 
in a small cone approximation, with the subsequent integrals evaluated numerically using a 
Monte Carlo approximation. The results were compared to more complete calculations~\cite{Ellis1} 
also using Monte Carlo evaluation of the integral, but without the small cone assumption \cite{Jager:2004jh}.

The code allows as input the colliding energy of the protons, the jet cone radius, the jet rapidity and 
the jet transverse momentum intervals for integration of the cross
section. The code requires assumptions for two scale inputs: the initial-state factorization 
scale ($\mu_{I}$) and the renormalization scale ($\mu_{R}$). We take the value 
$\mu_{I} = \mu_{R} = p_{T}$ of the jets. The polarized and unpolarized inclusive 
jet cross sections are calculated by separate programs.

As originally configured, the programs require polarized and unpolarized parton 
distribution functions that are sampled by the Monte Carlo portions of the program 
over a wide range of $x$ and $Q^{2}$. These are tabulated at fixed values beforehand 
and interpolated to the
required precision. The original configuration also allowed for a selection among 
several sets of unpolarized (CTEQ5, CTEQ6M,
CTEQ6M.1) and polarized (GRSV2000 STND, MAX, MIN, ZERO) parton distribution functions. All calculations 
here use the pdf set CTEQ6M \cite{CTEQ6M} and a cone radius of 0.4 or 0.7, as noted.

\begin{figure}
\includegraphics[width = 0.50\textwidth]{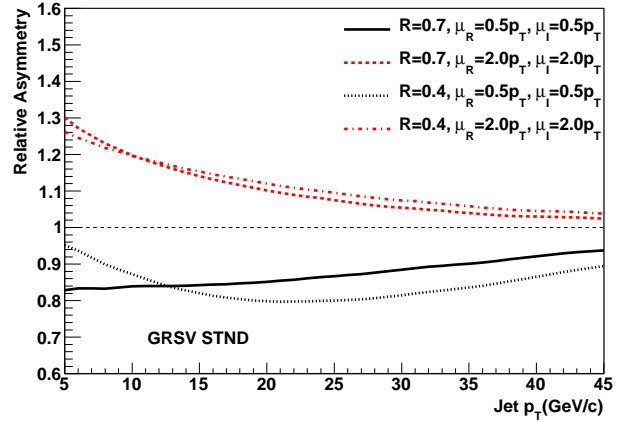}
\caption{\label{theory_1}  (color online) Scale dependence of the relative asymmetry
(defined in the text) as a function of jet transverse momentum. The asymmetry is 
calculated for the GRSV Standard set of parameters. The black (red) lines are the 
ratios of the asymmetries calculated for half
(twice) the nominal renormalization and initial state scales of $\mu_{R} = \mu_{I} =
p_{T}$. The dotted curves are for a jet cone radius of 0.4 and solid and dashed curves
are for a cone radius of 0.7.}
\end{figure}

%As pointed out by Ellis and Soper \cite{Ellis2}, the accuracy of these calculations 
%with respect to order $\alpha_{s}$ and scale are related by the renormalization group 
%equations. 

%The possible size of NNLO corrections to the cross sections are expected to be at least as large 
%as the variation of the cross section with respect to the scale factors.

The possible size of higher order corrections to the cross sections is conventionally
estimated by varying the factorization and renormalization scales by a factor of 2 about
the nominal scale. 
In Fig.~\ref{theory_1}, we plot calculations that
show the scale-dependence of the relative asymmetry, $A_{LL}/A_{LL}^{0}$, 
%show the scale-dependence of the relative asymmetry, $(A_{LL} - A_{LL}^{0})/A_{LL}^{0}$, 
as a function of jet transverse momentum
for both jet cone radii. In this expression, $A_{LL}^0$ is the asymmetry calculated with 
the nominal scales $\mu_{I,R} = p_{T}$, while
$A_{LL}$ is calculated for scales of $2p_{T}$ and $p_{T}/2$. The parton distribution functions 
used in this calculation were the CTEQ6 set and GRSV2000 STND.
In general, the larger the cone radius, the less sensitive is the calculation of the spin asymmetry 
(not cross section) with respect to scale and higher order contributions.
The choice of cone radius equal to 0.4 for 2005 was made 
in consideration of the size of the (instrumented) part
of the detector, in order that acceptance uncertainties would not dominate the systematic 
uncertainty on the jet energy.
With increased EMC coverage in year 2006, our sensitivity to scale variations is lessened.

The impact of these data on previous constraints of the integral of the polarized gluon
distribution function, $\Delta G$, from deep-inelastic
lepton scattering data is evident in Figs. \ref{figure_13} and \ref{figure_14}.
The measured $A_{LL}$ values are seen to lie in the region below the previous best-fit DIS curve,
GRSV standard \cite{Gluck:2000dy,Jager:2004jh,SVpriv}. The remaining curves associated with GRSV
are the polarized parton distributions re-fit to constrain $\Delta{G}$ to a series of values
spanning the full range $-g(x) \leq \Delta{g}(x,Q^2_0) \leq g(x)$, that is the gluon spins may be fully
polarized in either direction, or interpolated to intermediate values using a common functional form.

Data from a single experiment cover a limited kinematic range in $x$ and $Q^2$,
making the measurement of the total integral $\Delta{G}(Q^2)$ at a specific $Q^2$
impossible. A rigorous extraction of $\Delta{G}$ requires the incorporation of these
inclusive jet asymmetries, along with other RHIC, DIS and SIDIS polarized scattering data,
into a global analysis. For example, the AAC analysis~\cite{aac} demonstrated that while the 
inclusion of PHENIX pion longitudinal double-spin asymmetries~\cite{phenix_data} available
at that time had only a small influence on the optimum fit, the uncertainties on the gluon 
polarized parton distribution function were significantly reduced
over the fit obtained using DIS data alone.

This type of analysis, more recently performed by de~Florian
\textit{et al}.\@ (DSSV) \cite{dssv,deFlorian:2009vb}, uses NLO pQCD fits to the world data set 
(including the STAR 2005 \cite{Abelev:2007vt} and a
preliminary version of the 2006 jet asymmetries presented here), constrained by a functional 
form that defines $\Delta{g}(x,Q^2)$ in the unmeasured regions of $x$ space,
to extract the spin-dependent parton densities. The DSSV global analysis is based on
Mellin moments, which allows a certain amount of $x$ integrated data to be included in the
fits. This is an especially important development for RHIC data, where the statistical precision
within our kinematical constraints 
thus far only allows the examination of $\Delta G$ over a limited range of $0.02 < x < 0.2$.
The DSSV best fit finds the gluon polarization to be much smaller than that in GRSV standard 
throughout the $x$ region which is currently constrained by data.  Furthermore, in the $x$ 
region sampled by RHIC data the DSSV $\chi^2$ + 2\% upper limit on $\Delta{g}(x,Q^2)$ 
at $Q^2$ = 10 GeV$^2$ is roughly half the GRSV standard value (see Fig.~2 in
\cite{dssv}).
%, clearly ruling out the previous GRSV best fit value.

\section{Summary}
In summary, we have reported an analysis of spin dependencies in the inclusive production
of mid-rapidity jets with transverse momenta up to $35\,\mathrm{GeV}/c$ in polarized $p$+$p$ collisions
at $\sqrt{s} = 200\,\mathrm{GeV}$ from data recorded in 2005 and 2006.
No evidence is found for the existence of statistically significant transverse asymmetries $A_N$, $A_{\Sigma}$, and $A_{TT}$. The $A_N$ result may provide new limits on the gluon Sivers distribution in the proton.
The longitudinal double-spin asymmetry $A_{LL}$ has been compared with NLO perturbative QCD evaluations based
on selected polarized parton distribution functions to demonstrate its sensitivity to the value of the gluon
helicity distribution inside the proton.

\section{Acknowledgments}
The authors thank W.~Vogelsang and M.~Stratmann for providing calculations and discussion.
We thank the RHIC Operations Group and RCF at BNL, and the NERSC Center at LBNL for their support.
This work was supported in part by the Offices of NP and HEP within the U.S. DOE Office
of Science; the U.S. NSF; the BMBF of Germany; CNRS/IN2P3, RA, RPL, and
EMN of France; EPSRC of the United Kingdom; FAPESP of Brazil;
the Russian Ministry of Education and Science; the Ministry of
Education and the NNSFC of China; IRP and GA of the Czech Republic,
FOM of the Netherlands, DAE, DST, and CSIR of the Government
of India; Swiss NSF; the Polish State Committee for Scientific
Research; SRDA of Slovakia, and the Korea Sci. \& Eng. Foundation.
Finally, we gratefully acknowledge a sponsored research grant for the 2006 run period from Renaissance Technologies Corporation.

\bibliography{basename of .bib file}

\end{document}